# Reducing the Volatility of Cryptocurrencies -- A Survey of Stablecoins

Ayten Kahya*, Bhaskar Krishnamachari*, Seokgu Yun[o]

*Viterbi School of Engineering, University of Southern California

[o]SovereignWallet Network Pte. Ltd.

## Abstract

In the wake of financial crises, stablecoins are gaining adoption among digital currencies. We discuss how stablecoins help reduce the volatility of cryptocurrencies by surveying different types of stablecoins and their stability mechanisms. We classify different approaches to stablecoins in three main categories i) fiat or asset backed, ii) crypto-collateralized and iii) algorithmic stablecoins, giving examples of concrete projects in each class. We assess the relative tradeoffs between the different approaches. We also discuss challenges associated with the future of stablecoins and their adoption, their adoption and point out future research directions.

## Introduction

The introduction of Bitcoin in 2009 revolutionized the world of finance by offering the first truly decentralized peer-to-peer protocol for digital cash. However, even as Bitcoin has been growing in popularity, spawning many other cryptocurrencies in its wake, their use as a medium of exchange has been challenging because they show high volatility, fluctuating greatly in price on a monthly, weekly, daily, sometimes even hourly basis. To address these challenges, researchers and developers have started to focus on the design of "stablecoins."

A stablecoin is a digital token on a blockchain that is designed to minimize price volatility with respect to a stable fiat currency or asset. The majority of stablecoins are pegged to fiat currencies such as USD, followed by assets such as gold or a basket of assets. This allows stablecoins to be utilized as primarily a unit of exchange as well as a unit of account and a store of value (if the underlying asset maintains value in the long term) compared to highly speculative volatile cryptocurrencies. Stablecoins are currently used for payments, trading, lending, investing, remittances and purchases. Volatility of cryptocurrencies worldwide in recent years enabled stablecoins to gain popularity across users and increase competition in financial markets. Indeed, in 2020 stablecoins have shown dramatic growth as various platforms experienced exponential growth in stablecoins use.[25]

Stablecoins can be designed in various ways depending on the desired utility. They are commonly useful for retail payments, international money transfer, while some stablecoins are designed for settlements between banks or sustaining an ecosystem around an activity. Depending on the design, stablecoins can increase efficiency of payments.[18] Stablecoins can be classified in three main categories as i) fiat or asset backed, ii) crypto-collateralized and iii) algorithmically stabilized stablecoins. There are also hybrid approaches, which may involve more than one type of backing such ad crypto and fiat backing. The degree of automation and centralization varies across the stablecoin types and use cases. Meanwhile, stablecoins market share grew during the impact of COVID-19 to global markets and cryptocurrency market crash following Bitcoin's large drop on March 12th 2020.[3] Investors turned to stablecoins amidst the market turmoil as the combined transfer of all stablecoins tracked by Coin Metrics reached $444.21M on March 13th. In January 2021, the US Office of Comptroller of Currency allowed national banks and federal savings associations to use stablecoins for bank-permissible functions.[31]

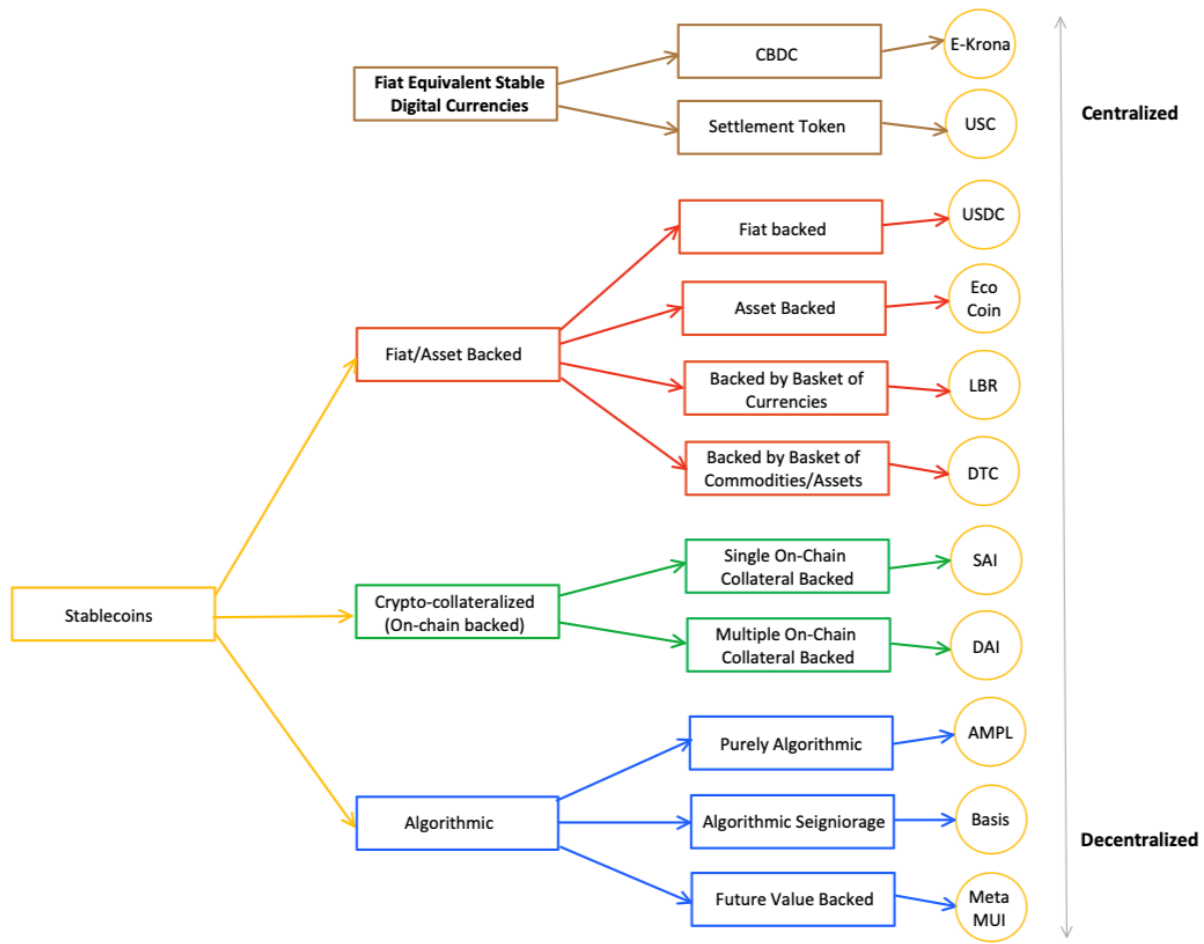

**Figure 1:** Stablecoins Categorization

## Types of Stablecoins

We classify different types of stablecoins based on how they try to stabilize the price. The first class we consider are fiat or asset-backed stablecoins, which directly connect the number of coins in circulation with either fiat currency or assets being held in reserve by some entity. We then discuss crypto-collateralized stablecoins, in which the asset being collateralized is itself a (potentially volatile) cryptocurrency. Finally, we discuss algorithmic stablecoins, which aim to utilize sophisticated smart contracts driven by external price-feeds to automate the process of minting and withdrawing coins from circulation to stabilize the price.

## Fiat or Asset Backed Stablecoins

Fiat or asset backed stablecoins are generally pegged to and backed one to one by an asset that is held in a reserve by a private bank. The most common type is USD pegged fiat backed stablecoins such as Tether, which has the highest market capitalization among all stablecoins.[9] Other approaches include backing by traditional assets such as gold (e.g. PAXG) or a basket of currencies such as one of the options considered in Diem (formerly called Libra, this project also includes individual currency-backed stablecoins). Although top ranked fiat backed stablecoin projects are generally able minimize volatility in regards to the peg, they can be impacted by fluctuations in the underlying asset's value. In the following we describe five types of asset/fiat backed stablecoins through exemplary projects.

**Single Fiat Backed - USDC**

Launched in 2018, USD Coin (USDC) is one of top fiat backed stablecoins[9]. USDC is an ERC20 token (also a token on the Stellar blockchain network) pegged to the USD.[8] Unlike decentralized stablecoins such as Dai, USDC is centralized as the token is issued by Centre, a consortium that is founded by the companies Coinbase and Circle.[7] It features multi-issuer scheme where eligible financial institutions need to meet various requirements such as being Anti Money Laundering (AML) audited and compliance with FATF standards. Centre aims to achieve a broad ecosystem where USDC can be integrated into other services and apps with this membership scheme and open-source framework. USDC can be used for trading, payments, cross border transactions, lending and investment.

Centre guarantees that each USDC is backed by one USD held in reserve by regulated institutions and they are always redeemable.[8] A significant feature claimed by Center about USDC is that it is audited by a well-known independent organization (Grant Thornton LLP) and regulated in the US.[7]

Coins are issued when a user requests USDC in exchange for USD. After the user transfers funds for tokens, CENTRE network verifies, mints and validates the USDC tokens to be transferred to the user. When a user requests a redemption, the network verifies and validates then removes the USDC by burning the tokens and returning the backing fiat to the user. Stability is ensured through this minting and burning process and the one-to-one USD backing of the coins.

**Stablecoins Backed by Basket of Fiat Currencies - Diem**

Formerly known as Libra (LBR), Diem is undoubtedly one of the most well-known stablecoin projects. It is developed by the Diem Association based in Switzerland (previously called the Libra Association), co-founded by the social networking giant Facebook.[10] Libra was planned to be backed by and pegged to a basket of assets initially consisting of USD, GBP, EUR and JPY with the Libra association governing the Libra network and managing the Libra Reserve for the backing assets.[22] According to its whitepaper "Libra's mission is to enable a simple global currency and financial infrastructure that empowers billions of people".[10] Due to the large outreach of Facebook and other well-established founding members including financial services and payments platforms companies, Libra was believed to emerge as a significant competitor in the global financial markets. Regulatory uncertainty of Libra's classification (as currency/derivative/ security/ commodity pool) in addition to other regulatory concerns such as fraud prevention received regulatory backlash and led to some of the big co-founders of the association such as Visa, Mastercard and eBay dropping out.[38] Consequently, the project has made various changes to the whitepaper and plans to launch at the end of 2020.[21] The novel version of the whitepaper states that Libra's vision has been to complement the fiat currencies rather than competing with them.[10] Concerns rose about Libra having potential to interfere with monetary policy and sovereignty rose if it were to scale up significantly and large volume of payments were to be made in LBR. Some viewed it as having potential to reduce reliance on a single-currencies for international trade (un-dollarization).[22] Hence, Libra moved forward with adding single-currency stablecoins from their proposed currency basket (e.g., LibraUSD, LibraEUR, LibraGBP or LibraSGD) to their platform, which will be fully backed by the Reserve.[10] While LBR will not be a separate asset from the single-currency stablecoins, it will rather be a digital composite of single-currency stablecoins defined with reference to fixed nominal weights. The team intends to work with regulators, financial institutions and central banks to increase the single currency stablecoins on their platform. Furthermore, Libra also aims to support the UN's Sustainable Development Goals.

**Asset Backed Stablecoins - Eco Coin**

ECO coin is a unique ecological cryptocurrency that is asset backed by trees.[35] Based on a circular economy concept, it acts an alternative digital currency that encourages environmentally sustainable actions through financial incentives.[20] The project was developed by Next Nature Network in Amsterdam and launched in 2017 with an implementation at a popular music festival.[36] Eco Coin is currently operating a pilot in a community.[20]

Users earn Eco Coins (ECOs) by sustainable actions at individual or organizational level; for instance, riding a bike to work or switching to green energy providers.[35] The value of these sustainable actions is determined according to their relative offsets of Carbon Dioxide emissions equivalent.[20] Additionally, ECOs can also be obtained via verifying sustainable actions as ECO inspectors, backing the token by contributing trees to the system, being a certified vendor or being a part of the technical development.[35] They can also be bought at the Initial Coin Offering. ECO inspectors, certified vendors and sensor integrated systems verify sustainable actions to prevent malicious actors from gaming the system. The platform is governed via the Decentralized Autonomous Charity (DAC) where ECO holders and stakeholders participate (by running a node) in votes for the development and decisions regarding ECOs.

Eco Coin has a unique issuance process. While each ECO is backed by a tree, one ECO coin is earned in exchange for contributing ten trees, while the other nine can be redeemed through sustainable actions. The trees are kept in escrow through the ECO Coin Foundation (ownership still belongs to the original owner). Since the lifespan of a tree is finite, the lifespan of an ECO is finite based on average tree lifespan. Thus, to make the system practical the coin deteriorates by a small percentage every year. If the average lifespan of the tree was 100 years, 1 ECO would represent a one-year-old tree and 0.01 ECO represent a hundred year old tree. A tree owner can only cut down and plant another as a replacement when the average tree lifespan is over.

After they are earned, ECOs can be spent in exchange for goods or services. The project aims to expand the sustainable marketplace where ECOs can be earned and used.[20] As the platform grows, the ECOs will be exchangeable with Euros according to the developers.

**Backed by a Basket of Assets/Commodities- Trade Coin**
Digital Trade Coin (DTC) is an example of asset backed stablecoin, which employs a unique approach to system design[1,2]. Currently in development at MIT, the project aims to explore an efficient and reliable digital currency that is trade-oriented, scalable, fast and environmentally friendly. DTC is pegged to real world assets such as energy, crops and minerals, which are supplied to the platform by a consortium of sponsors as reserve collaterals (backing) in exchange for DTC tokens. Sponsors may include alliances of small nations, commercial trades, business or farmers etc. In the DTC ecosystem, DTCs are traded amongst sponsors while non-sponsors (users) can obtain "e-Cash" from the consortium that is backed by DTCs in exchange for fiat money. E-Cash serves a stable payment method for everyday transactions and can store value over time. It is important to note that financial transactions involving fiat currencies are carried out through a narrow bank. In addition to e-Cash, if a participant wants to obtain newly minted DTC; they transfer cash to the narrow bank, which transfers money to the sponsors who in turn release DTC to the participant. This way, the participant turns to a shareholder in the pool. To redeem fiat money, the participant can return the DTC to the administrator who sells assets to return the cash and burns the DTC.

The system is governed by the consortium and its delegated administration responsible with carrying out monetary policies of the consortium and controlling various system functions. Stability of the DTC is also maintained by the consortium. When the market price of DTC falls significantly below the market price of the relevant asset pool, economic agents will return DTC to the administrator. The administrator will sell the corresponding amount of assets to return the proceeds to these agents. Conversely, if the market price of DTC is significantly above the market price of the relevant asset pool, sponsors will contribute more assets to the pool. So, the administrator can issue more DTCs to sponsors that sell them to other participants for cash thereby pushing the DTC price down.

According to the developers of DTC, the complexity of system design and various system functions depend on specific applications of the concept. There are three layers of ledgers within the system architecture; recording of the assets is done through the Assets Ledger while the coin transactions are enabled through the Coins Ledger. Lastly, E-Cash transactions take place on the Transactions Ledger. Additionally, DTC ledger can be designed as semi-private to meet AML / Know Your Customer (KYC) standards. In order to establish a more efficient system and avoid energy intensive mining methods, DTC network will utilize a set of trusted nodes as validators. DTC concept is currently being explored through two pilot projects related to international commerce and commodity markets.

**Settlement Coin - Fnality**
Fnality, formerly known as Utility Settlement Coin (USC), utilizes a similar approach to fiat-backed stablecoins, albeit focused on the problem of bank settlements. The project is developed by a consortium of banks including some of the world's major banks and financial institutions.[15] Fnality aims to establish a decentralized Financial Market Infrastructure in each currency on its platform to deliver means of payment for wholesale banking markets via its tokenised settlement asset USC. USC will operate on a private ledger on an Enterprise Ethereum blockchain and based on the jurisdiction of the relevant central bank money, it will act as a digital representation of an entitlement, claim or interest.[14] USC will serve as a medium-of-exchange for the wholesale market and as a store of value meant solely to help settlement.

The primary distinction of USC from other coins is that the aforementioned digital representation is backed by corresponding assets at the respective central banks. The initial currencies on USC's platform will be CAD, EUR, GBP, JPY and USD while more currencies might be added in the future.[16] Furthermore, USC plans to be fully backed with guarantee of exchangeability into fiat currency anytime.  The key aspect of Fnality's innovation is the facilitated finality of settlements. Settlement is achieved in compliance with local settlement finality laws and regulations.  Thus, the finality and irreversibility (by court) of the settlement is ensured locally, for each jurisdiction.[14] Developers believe Fnality will reduce liquidity needs and facilitate cash management by removing the need of "having many separate accounts at Correspondents and Custodians".  This also reduces the settlement time and complexity by enabling international banks to easily transfer ownership of USC.[23]

## Crypto-Collateralized Stablecoins
Crypto-collateralized stablecoins (a.k.a. on-chain backed stablecoins) are backed by other cryptocurrencies on the blockchain. The core component is over-collateralizing the backing cryptocurrencies so that their volatilities have minimal impact on the stablecoin's price. However, they may be impacted by severe changes in collaterals' price. Various projects mitigate this problem by multiple on-chain asset backing to reduce the dependence on a single type of collateral. In this section we describe how collateralized stablecoins work through analyzing MakerDao.

**Maker Dao**

Launched in 2017 by MakerDAO, Dai is a crypto-collateralized token soft-pegged to the USD.[9] The Maker Protocol is amongst the largest de-fi applications on Ethereum as well as the leading crypto collateralized stablecoin by market capitalization. Dai has no fiat backing and there is no central authority in the Maker Protocol issuing the tokens.[28] It can be traded on various exchanges, used for payments and transactions, lent or held for savings via Dai Savings Rate (DSR). Although the Maker Foundation founded MakerDAO and bootstrapped the Maker Governance, they plan to dissolve once the DAO (Decentralized Autonomous Organization) is ready to fully govern the platform. The initial single collateral Dai on the platform (backed by ETH) was called 'Sai' after transitioning to the new Maker Protocol with multiple collateral types. Sai officially shutdown in May 2020.[26]

Dai is generated when a user locks in excess collateral in a "Vault".[28] During this process, the collateralization ratio needs to be set above the liquidation ratio at which collateral becomes too risky. Liquidation ratio is a key risk parameter that is determined by the governance according to the risk characteristics of collaterals that helps keep stability of the token. Maker Protocol currently accepts Ether (ETH), Basic Attention Token (BAT), USD Coin (USDC) and Wrapped Bitcoin (WBTC) tokens as collateral for Dai. MakerDAO community is also considering including tokenized trade invoices and music streaming loyalties as collaterals for Dai.[33] When the collateral debt is paid with the stability fee, the vault is then closed while the collateral is returned to the user and the corresponding Dai is burnt from supply. Stability fee acts as an interest rate and is one of the primary features of Dai's stability mechanism. Lower stability fee encourages users to open more Vaults and borrow Dai, thus increasing the Dai in circulation and lowering the price when Dai's market price is above the target price of 1 USD. Similarly, higher Stability Fee incentivizes users to close Vaults, thus removing Dai from circulation and increasing the price when Dai's market price is less than the target price. Each type of collateral has a specific stability fee determined by Maker Governance. DSR also helps maintain stability through active governance by MKR holders. When the market price of Dai is above the target price governance can vote to decrease DSR to reduce demand thereby reducing the price of Dai and vice versa.

Liquidation is a significant concern for Dai users and it also encourages them to help maintain stability. If a Vault becomes too risky, it is automatically liquidated and sold in internal market-based auction mechanisms starting with collateral auctions. The aim is to cover the vault obligations plus a liquidation penalty fee pertaining to the collateral type. When all the debt and fees are covered via the auction proceeds, the system returns remaining collateral to the user. If the auction falls short of covering the Vault obligations, the deficit becomes Protocol debt, which the system tries to recover first through a buffer and then a debt auction if there is remaining debt. Additionally, there are other mechanisms and external actors that help maintain stability of Dai such as multiple trusted Oracle Feeds resembling a decentralized oracle infrastructure and keepers that participate in Maker auctions.

Maker Governance Community is responsible to govern the protocol by managing the platform and associated financial risks. Any user on the platform can propose a change or an update to the system while only MKR holders can vote. A user's MKR holdings determine their voting power in proposals.

There are various incentives for governance to responsibly govern the protocol including the debt auction where MKR is minted and sold to recapitalize the system. Another highlight of the governance abilities is it can also protect the protocol from a malicious attack or long-lasting market irrationality by initiating an Emergency Shutdown as a last resort.

Dai (SAI) has shown resilience to fluctuations in ETH prior to 2020.[9] Meanwhile, during the crypto market collapse in March 2020, Dai faced a near death situation where many vaults became under-collateralized resulting in a large number of auctions. Some of these auctions were won by zero-bidders who bid decimal amounts, consequently there was a shortfall of more than 5.4 M DAI.[29] The system was recapitalized through debt auctions, where MKR was auctioned for Dai (reducing MKR value).

**Synthetic Assets**
Synthetic assets enable users to gain exposure to underlying assets without necessarily holding them.[24, 37] The leading example is the Synthetix protocol, which enables the issuance of synthetic assets called Synths on the Ethereum blockchain. The platform's native token SNX is used as collateral to mint Synths. Synthetic commodities that the platform supports range from cryptocurrencies, real-world assets such as gold, indexes and inverses.[24] These synthetic assets such as the synthetic USD (sUSD) or synthetic Ether (sETH) track the price of and hold a stable value with respect to the underlying asset (e.g. sUSD's price is around 1 USD).[9, 37] Similarly, wrapped coins such as WBTC (which is a collateral type for DAI) can be considered as examples of synthetic assets. Wrapped coins are non-native coins on a blockchain tied to the value of another cryptocurrency that originate from a different blockchain.[5] This functionality of usability on a different blockchain is achieved by putting the backing coin in a type of digital vault called wrapper. Although synthetic assets are not necessarily always stablecoins, it is important to note them in this context as they can be similar to crypto-collateralized or crypto-backed stablecoins.

## Algorithmically Stabilized Stablecoins

Algorithmic stablecoins do not essentially require the use of backing assets. Such coins typically solely depend on algorithmic stabilization, oracle price feeds and user participation (trading) to maintain their peg. Although a truly stable algorithmic coin remains to be achieved, there are an increasing number of projects in this area. In the following we describe several types of algorithmic stabilization via examples.

**Purely Algorithmic - Ampleforth**
Ampleforth is an example of a purely algorithmic approach to reducing the volatility of cryptocurrencies. It is a synthetic commodity money based on algorithmically enforced elastic supply.[13] The Ampleforth platform has a single ERC20 token called AMPL.[9] It should be noted that Ampleforth does not claim to be a stablecoin but rather a low volatility coin that is designed to diversify risk.[13, 30]

High correlation among cryptocurrencies results in a vulnerable ecosystem and introduces systemic risk. Ampleforth's elastic supply tackles this challenge by an algorithmic rebasing mechanism that applies countercyclical pressure against the fluctuation in the market. The rebasing mechanism helps maintain stability by incentivizing users to stabilize the system via arbitrage opportunities. If the market price of AMPL is above the Price Target plus the Price Threshold, then the algorithm expands the token supply, reducing the price. Whereas, if the market price of AMPL falls below the price target minus the price threshold, the algorithm contracts the supply by automatically and directly removing tokens in user accounts to increase the AMPL price. Moreover, the changes to algorithmically determined supply targets are graded over a defined time to distribute uniformly over this period. During the expansion phase, there is a limited sell opportunity for fast actors; while during a contraction phase there is limited buy opportunity. This buy and sell opportunity incentivizes traders to correct the price and bring the system to equilibrium after expansion or contraction phases. As long as enough traders are willing to benefit from trading opportunities, the platform can be maintained theoretically.

**Algorithmic Seigniorage - Basis**

Despite the failed launch of the token, the algorithmic seigniorage design of the Basis token is noteworthy.[19] The protocol was designed to maintain stability by expanding and contracting the supply when the market price of the token deviates from the peg, while price information would be provided by oracles. It featured a three-token system, with Basis as the stablecoin pegged to the USD, bond and share tokens.[34] Basis and bond tokens would be issued, share tokens would have a fixed supply at genesis and return Basis to shareholders when the platform expanded. Basis failed to launch due to regulatory reasons associated with bond and share tokens and the funds were returned to investors.[32]

If Basis would trade for less than one USD, then the protocol would issue and auction Bond tokens to users in order to remove coins out of circulation to increase token's price. The auctions would run continuously until enough Basis is destroyed. Bond tokens would be auctioned to contract Basis supply and then buyers would be able to redeem one Basis in the future for a price of less than one Basis at the time of the auction. Conversely, if Basis traded for more than one USD, new Basis would be issued to increase supply and lower the price towards the peg. During the expansion cycle, Bond holders would receive newly minted Basis tokens with oldest bonds first in queue order. However, bonds older than 5 years would be expired to prevent bonds from losing value. After the outstanding bonds are covered, remaining newly minted Basis would be evenly distributed across share tokens. Basis developers expected only small volatility as long as there is enough liquidity and speculators incentivized to participate in auctions to restore the peg.

**Future Value Backed - MetaMUI**

MetaMUI is the mainnet coin of MUI MetaBlockchain, a digital currency generation platform developed by Sovereign Wallet Network.[40] The value of MetaMUI coin (digital sovereign currency) is controlled and maintained algorithmically by a special AI-based algorithmic engine called ACB (Algorithmic central bank). In contrast to other stablecoins described in this survey which not only reduce volatility but also typically maintain a constant value with respect to a fiat currency, the MetaMUI coin is designed and developed to maintain low volatility while increasing value over long periods of time.

For publishing a digital currency, the central bank node of the target currency is required to deposit assets to MUI central bank node. And these assets, called project funds, are collected and maintained by the MUI treasury. The project funds are used to buy MetaMUI coins from the market and increase its demand. Once the demand is high, the MUI ACB sells the coins at a higher value in the market and increases the circulation up to a capped limit.

The most important goals of the MetaMUI coins are to maintain an increasing value over time with respect to fiat currencies and maintain low volatility within any short window of time. To achieve this, MUI ACB calculates next target price based on two prices- previous market price of MetaMUI and leveraged market price of gold and other assets that are appreciating over time with respect to fiat currencies.

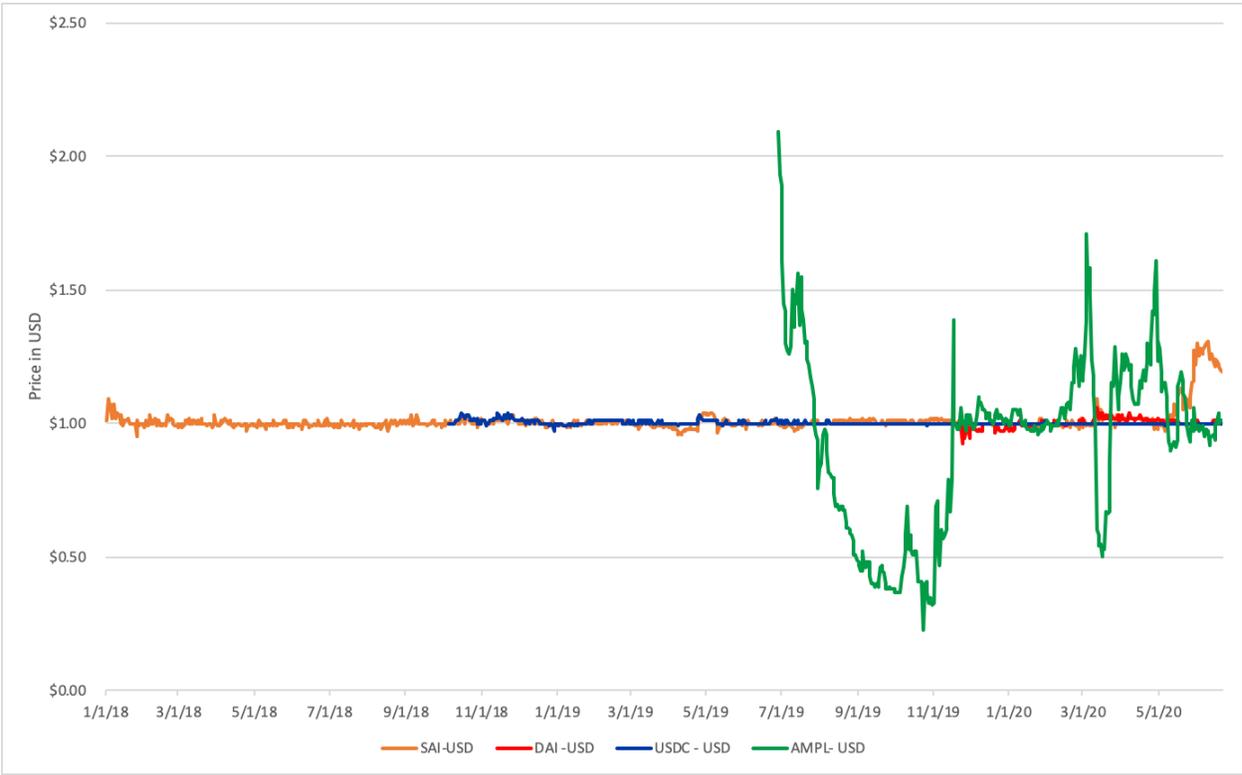

**Figure 2:** Price (USD) chart for USDC, DAI, SAI and AMPL (source: https://www.coingecko.com/en)

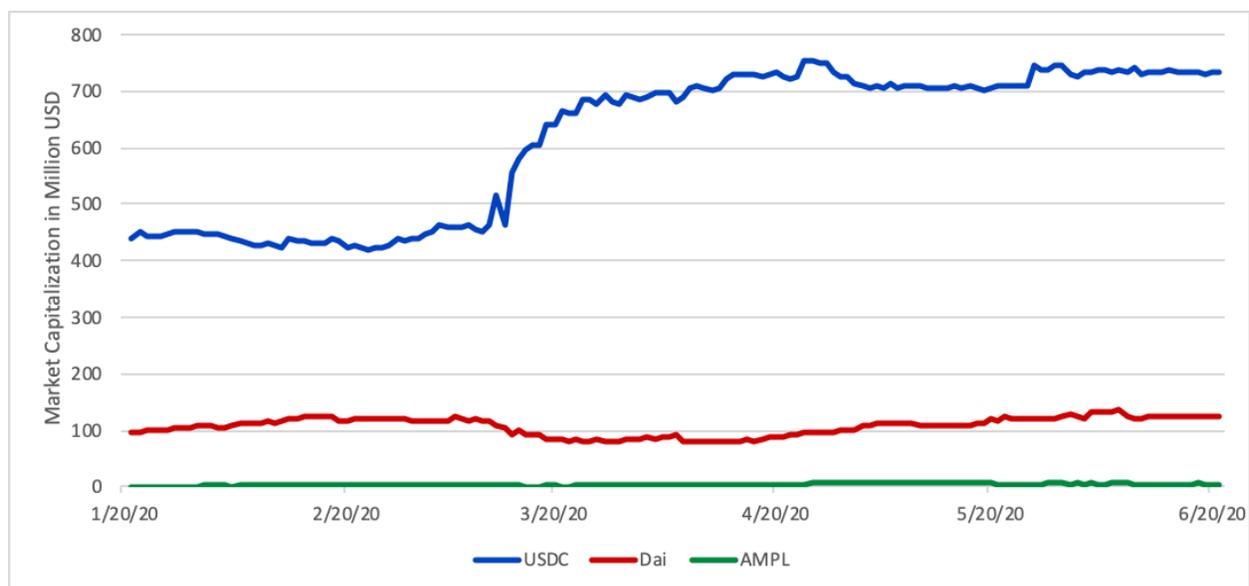

**Figure 3:** Market Capitalization in USD for USDC, DAI and AMPL (source:https://coinmarketcap.com/)

## Central Bank Digital Currencies (CBDCs)

Following Bitcoin, the rising popularity of stablecoins such as Libra intrigued central banks across the world to explore blockchain and another form of stable digital currency, namely Central Bank Digital Currencies (CBDC). In fact, several central banks across the world are exploring CBDC projects such as the Sveriges Riksbank (Sweden's e-krona), the Chinese Central Bank (digital yuan), the Eastern Caribbean Central Bank and the Central Bank of Brazil while other central banks such as the Bank of England and the Bank of Canada are considering CBDCs with ongoing research.[39, 44] CBDC can be thought of as a digital money equivalent to physical cash or reserves held at a central bank.[45] Depending on the particular scheme or use case, a CBDC design may or may not include a blockchain DLT. However due to the benefits of DLT, blockchain-based CBDCs are widely considered. There is ongoing exploration of the role that commercial banks would play with respect to CBDC's.[17]

Major benefits of DLT/blockchain based CBDCs include faster and cheaper domestic or cross-border payments with respect to traditional payment methods, reduced friction associated with traditional banking, resilience against operational failures, physical disruptions and cyberattacks that traditional systems are vulnerable to.[44] Although there are substantial perceived benefits to implementing CBDC, regulatory aspects, privacy concerns, challenges associated with the scalability of blockchain technology, energy consumption and negative impacts on the financial system (more specifically commercial banks and fractional banking) need further assessment and development.[44, 18]

## Challenges & Risks Regarding Stablecoins

Despite the advantages of stablecoins, there are legal, regulatory and oversight risks and challenges associated with stablecoins. Firstly, from a legal perspective, the categorization of stablecoins is relatively ambiguous. Depending on the jurisdiction and the characteristics of a stablecoin, it may be considered an equivalent to money, a contractual claim, implicating a right against underlying assets, a security or a financial instrument.[18] It is hard to regulate stablecoins without legal certainty. This uncertainty also complicates consumer and investor protection where adequate information and disclosures including the risks and obligations should be available for customers or investors to make informed decisions. Additionally, due to the lack of proper supervision and effective regulations, stablecoins can be potentially used for illicit financial activities, money laundering and financing terrorism, thereby compromising financial integrity.[41] To mitigate this problem, entities and issuers in a stablecoin system should comply with the highest international AML/KYC and countering the financing of the proliferation of weapons of mass destruction (CPF) standards.[18] Overall, the challenges and risks associated with each stablecoin depends on its design and structure as well as the jurisdiction that it is in. While some risks such as money laundering might be easier to address for certain types of stablecoins in certain jurisdictions, others might be more complicated. Additionally, the distributed nature of blockchain networks may make it difficult to enforce regulations such as tax compliance.

Depending on the backing of the stablecoin and how they are held, stablecoins might not be able to maintain stability and redeemability/convertibility with respect to the peg.[11, 4] Lack of transparency regarding collateral backing of some well-known stablecoins such as Tether has received scrutiny over the recent years.[46] This accelerated regulatory compliance efforts among the fiat backed stablecoin space. Likewise, large price fluctuations can pose financial or operational failure risks for the stablecoin as a payment system. Moreover, poorly designed or governed systems introduce systemic risk and pose disruptions to financial markets and the economy.[18] This risk is amplified for stablecoins that are adopted at a larger scale. Additionally, lack of interoperability among stablecoins and other payment systems can lead to inefficiencies and isolated financial silos.[4] Wrapped tokens and the use of atomic swaps can potentially improve interoperability between blockchains.

DLT benefits from eliminating risks from a single point of failure. They have resiliency benefits against various cyber and operational risks compared to centralized systems.[12] On the other hand, cyber security is still a significant concern for DLT systems.[43, 44] Operational resilience for a stablecoin is also essential as it can be compromised by black swan events, malicious attacks to the system or severe market downturns. Eventually, stablecoins might be subject to international standards such as ISO or IEC standards and regulations regarding operational and cyber risks.[18] Furthermore, holders may lose confidence in the stablecoin if the issuing organization or governance is not stable and accountable compared to central and commercial banks. As an emerging technology DLT can also be susceptible to currently unknown risks.[44]

Ensuring market integrity (fairness and transparency of the price information) is another challenge pertinent to stablecoins that needs to be addressed by maintaining fair and stable prices in primary and secondary markets.[18] Entities taking on multiple roles such as trading platform, market-maker and custodial wallet might increase the risk of market misconduct due to conflicts of interest.

Stablecoins implemented on top of open, permissionless DLT platforms inherit some of the fundamental challenges associated with DLT protocols. These include high energy consumption with Proof of Work (which could be mitigated by alternative approaches under development such as Proof of Stake), interoperability as well as the problem of low transaction throughput.[43] On the privacy side, there are concerns about data collection and usage that might discourage users from using stablecoins.[41] Since blockchain serves as an immutable distributed ledger, it conflicts with a legal right in various jurisdictions, the "right to be forgotten".[43] Thus, organizations need to carefully evaluate user's right to privacy especially with public blockchains in such jurisdictions. It should be mentioned that there are a number of efforts in the cryptocurrency space e.g. Monero, Zcash, Aztec and Nightfall focused on privacy using zero-knowledge proofs, which could be applied or extended to provide privacy guarantees for stablecoin transactions as well, though a challenge is that these schemes have to be compliant with AML laws as well.[42]

In addition to the general risks and challenges discussed above, there are various distinct challenges that apply to the particular categories of stablecoins. Contrary to the idea of decentralization, fiat backed stablecoins are relatively centralized as they require a trusted institution or a consortium to issue, burn or hold assets. Due to backing requirements associated with handling assets, operations can be more costly compared to other stablecoins. Meanwhile, they may be less complex in design and less volatile than most crypto collateralized or algorithmic stablecoins.

In the case of crypto-collateralized stablecoins, loans may not be fully recovered in the case of default due to high fluctuations in the collateral's value. Additionally, tokens with multiple on chain collaterals incur a correlation risk, which implies the diversification benefit will be less if the collaterals' volatilities have high correlation.[27] Increasing exposure in one type of collateral can impose similar risks. Low quality price feeds (often delivered through centralized oracles) are a significant source of risk relevant to crypto-collateralized and algorithmic stablecoins that can adversely affect stability and operational resilience. Crypto-collateralized stablecoins also need more careful design due to the possibility of liquidity issues and need to account for human factors with respect to incentives for opening/closing collateralized deposits.

Algorithmic stablecoins are highly complex; issuance and stability factors might not be fully understandable for users. Since they do not feature any collateral, pure-algorithmic stablecoins are most vulnerable to market crashes and "death spirals".[19] Additionally, bond or share tokens in algorithmic seigniorage may be classified as securities in some jurisdictions as users can make profit through them. Algorithmic stablecoins depend more heavily on buy and sell activity of users with rational economic incentives to maintain stability but if the participants lose interest in buying and selling, the peg cannot be maintained.

The total market capitalization of stablecoins worldwide has recently reached approximately 10 Billion USD, which is still relatively extremely small compared to all the fiat money in the world.[47] Many challenges and risks as aforementioned need to be tackled for stablecoins to be adopted on a global scale. According to the Bank of England, efficiency and functionality benefits over current payment systems are needed for wider adoption of stablecoins.[4] If stablecoins achieve global scale, they could present risks and challenges to monetary policy, international monetary system, financial stability and fair competition.[18] Future research is needed to determine at what level stablecoin usage could present a risk to implementation of monetary policy.[6] Many countries and well-known financial institutions are researching and/or developing CBDCs to reap the benefits of blockchain based stable digital currencies while avoiding the potential risks and adverse impacts of stablecoins. CBDCs offer a more scalable, secure and stable digital currency depending on their design and structure. The concept is still mostly in an experimental stage, and thus further research is required to assess the financial impacts of CBDCs.

## Conclusions

We have presented a survey classifying and describing the many different kinds of stablecoins that are being researched and developed, including fiat/asset-backed stablecoins, crypto-collateralized stablecoins, and algorithmic stablecoins, as well as the closely related efforts on developing fiat-equivalent digital currencies. While some projects are further along in deployment, all projects are at a relatively early stage with a lot of open questions, particularly related to risk management. We believe that significant new research is needed to address these challenges.